\documentclass[12pt]{article}

\begin{document}

\author{C. Bizdadea\thanks{%
e-mail address: bizdadea@central.ucv.ro}, C. C. Ciob\^{\i }rc\u {a}\thanks{%
e-mail address: ciobarca@central.ucv.ro}, E. M. Cioroianu\thanks{%
e-mail address: manache@central.ucv.ro}, \and S. O. Saliu\thanks{%
e-mail address: osaliu@central.ucv.ro}, S. C. S\u {a}raru\thanks{%
e-mail address: scsararu@central.ucv.ro} \\
Faculty of Physics, University of Craiova\\
13 A. I. Cuza Str., Craiova RO-1100, Romania}
\title{Interacting mixed-symmetry type tensor gauge fields of degrees two and
three: a four-dimensional cohomological approach }
\maketitle

\begin{abstract}
A special class of mixed-symmetry type tensor gauge fields of degrees two
and three in four dimensions is investigated from the perspective of the
Lagrangian deformation procedure based on cohomological BRST techniques. It
is shown that the deformed solution to the master equation can be taken to
be nonvanishing only at the first order in the coupling constant. As a
consequence, we deduce an interacting model with deformed gauge
transformations, an open gauge algebra and undeformed reducibility
functions. The resulting coupled Lagrangian action contains a quartic vertex
and some ``mass'' terms involving only the tensor of degree two. We discuss
in what sense the results of the deformation procedure derived here are
complementary to recent others.

\end{abstract}

The problem of constructing consistent interactions that can be introduced
among gauge fields in such a way to preserve the number of gauge symmetries
\cite{alpha1}--\cite{alpha3} has been reformulated as a deformation problem
of the master equation \cite{def} in the framework of the antifield BRST
formalism \cite{1}--\cite{5}.

In this paper we generate all consistent interactions that can be added to a
special class of mixed-symmetry type tensor gauge fields, described in the
free limit by the Lagrangian action
\begin{equation}
S_{0}^{L}\left[ A_{\alpha (\lambda )},B^{\alpha \beta (\lambda )}\right]
=\int d^{4}x\partial _{\left[ \alpha \right. }A_{\left. \beta \right]
(\lambda )}B^{\alpha \beta (\lambda )},  \label{cin1}
\end{equation}
where the tensor gauge field of degree three is assumed antisymmetric in its
first two indices, $B^{\alpha \beta (\lambda )}=-B^{\beta \alpha (\lambda )}$%
, while that of degree two displays no symmetry. We work with the
conventions that the Minkowskian metric $g_{\mu \nu }$ is of
`mostly minus' signature $\left( +,-,-,-\right) $, and that the
completely antisymmetric four-dimensional symbol $\varepsilon
^{\alpha \beta \gamma \delta }$ is valued like $\varepsilon
^{0123}=+1$. On the one hand, models with mixed-symmetry type
tensor gauge fields attracted much interest lately. In this
context, more problems, like, for instance, the interpretation of
the construction of the Pauli-Fierz theory \cite{pf}, the dual
formulation of linearized gravity \cite{dual}--\cite{lingr}, the
impossibility of consistent cross-interactions in the dual
formulation of linearized gravity \cite{lingr},  or the general
scheme for dualizing higher-spin gauge fields in arbitrary
irreducible representations of $GL(D,\mathbf{R})$ \cite{dualsp},
have been reanalyzed. On the other hand, the action (\ref{cin1}%
) can be regarded in some sense as a topological Batalin-Fradkin
(BF) theory. It is known that BF-like theories are deeply
connected with two-dimensional gravity \cite{gr1}--\cite{gr7} via
the so-called Poisson Sigma Models \cite{psm1}--\cite{psm7}. In
view of its links with some important classes of gauge theories,
we believe that the study of the theory under consideration might
bring some contributions to the quantization of gravity without
string theory.

Action (\ref{cin1}) is found invariant under the abelian gauge
transformations
\begin{equation}
\delta _{\epsilon }A_{\alpha (\lambda )}=\partial _{\alpha }\epsilon
_{(\lambda )},\;\delta _{\epsilon }B^{\alpha \beta (\lambda )}=\varepsilon
^{\alpha \beta \gamma \delta }\partial _{\gamma }\epsilon _{\delta
}^{\;\;(\lambda )},  \label{cin2}
\end{equation}
where all gauge parameters are bosonic. The gauge generators of the tensor
fields $B^{\alpha \beta (\lambda )}$ are off-shell first-stage reducible
since if we make the transformation $\epsilon _{\delta }^{\;\;(\lambda
)}=\partial _{\delta }\theta ^{(\lambda )}$, with $\theta ^{(\lambda )}$
arbitrary functions, then $\delta _{\epsilon }B^{\alpha \beta (\lambda )}=0$%
. In consequence, we deal with a free normal gauge theory, of Cauchy order
three.

In order to investigate the problem under consideration, we employ the
antifield-BRST formalism. The BRST complex includes, besides the original
tensor fields, the fermionic ghosts $\left( C_{(\lambda )},\eta _{\alpha
}^{\;\;(\lambda )}\right) $ respectively associated with the gauge
parameters $\left( \epsilon _{(\lambda )},\epsilon _{\alpha }^{\;\;(\lambda
)}\right) $, the bosonic ghosts for ghosts $\eta ^{(\lambda )}$ due to the
first-stage reducibility relations, as well as their antifields, denoted as
star variables. The BRST differential for this free model ($s$) decomposes
as the sum between the Koszul-Tate differential and the exterior
longitudinal derivative only, $s=\delta +\gamma $. The Koszul-Tate complex
is graded in terms of the antighost number ($\mathrm{agh}$), such that $%
\mathrm{agh}\left( \delta \right) =-1$, $\mathrm{agh}\left( \gamma \right) =0
$, while the degree of the exterior longitudinal complex is known as the
pure ghost number ($\mathrm{pgh}$), with $\mathrm{pgh}\left( \gamma \right)
=1$, $\mathrm{pgh}\left( \delta \right) =0$. The degrees of the BRST
generators are valued like
\begin{equation}
\mathrm{pgh}\left( B^{\alpha \beta (\lambda )}\right) =\mathrm{pgh}\left(
B_{\alpha \beta (\lambda )}^{*}\right) =\mathrm{pgh}\left( A_{\alpha
(\lambda )}\right) =\mathrm{pgh}\left( A^{*\alpha (\lambda )}\right) =0,
\label{cin4}
\end{equation}
\begin{equation}
\mathrm{pgh}\left( \eta _{\;\;\;(\lambda )}^{*\alpha }\right) =\mathrm{pgh}%
\left( C^{*(\lambda )}\right) =\mathrm{pgh}\left( \eta _{(\lambda
)}^{*}\right) =0,  \label{cin5}
\end{equation}
\begin{equation}
\mathrm{pgh}\left( \eta _{\alpha }^{\;\;(\lambda )}\right) =\mathrm{pgh}%
\left( C_{(\lambda )}\right) =1,\;\mathrm{pgh}\left( \eta ^{(\lambda
)}\right) =2,  \label{cin6}
\end{equation}
\begin{equation}
\mathrm{agh}\left( B^{\alpha \beta (\lambda )}\right) =\mathrm{agh}\left(
A_{\alpha (\lambda )}\right) =\mathrm{agh}\left( \eta _{\alpha
}^{\;\;(\lambda )}\right) =\mathrm{agh}\left( C_{(\lambda )}\right) =0,
\label{cin7}
\end{equation}
\begin{equation}
\mathrm{agh}\left( \eta ^{(\lambda )}\right) =0,\;\mathrm{agh}\left(
B_{\alpha \beta (\lambda )}^{*}\right) =\mathrm{agh}\left( A^{*\alpha
(\lambda )}\right) =1,  \label{cin8}
\end{equation}
\begin{equation}
\mathrm{agh}\left( \eta _{\;\;\;(\lambda )}^{*\alpha }\right) =\mathrm{agh}%
\left( C^{*(\lambda )}\right) =2,\;\mathrm{agh}\left( \eta _{(\lambda
)}^{*}\right) =3,  \label{cin9}
\end{equation}
while the operators $\delta $ and $\gamma $ act on them via the definitions
\begin{equation}
\delta B^{\alpha \beta (\lambda )}=\delta A_{\alpha (\lambda )}=\delta \eta
_{\alpha }^{\;\;(\lambda )}=\delta C_{(\lambda )}=\delta \eta ^{(\lambda
)}=0,  \label{cin10}
\end{equation}
\begin{equation}
\delta B_{\alpha \beta (\lambda )}^{*}=-\partial _{\left[ \alpha \right.
}A_{\left. \beta \right] (\lambda )},\;\delta A^{*\alpha (\lambda
)}=2\partial _{\beta }B^{\beta \alpha (\lambda )},  \label{cin11}
\end{equation}
\begin{equation}
\delta \eta _{\;\;\;(\lambda )}^{*\alpha }=\varepsilon ^{\alpha \beta \gamma
\delta }\partial _{\beta }B_{\gamma \delta (\lambda )}^{*},\;\delta
C^{*(\lambda )}=-\partial _{\alpha }A^{*\alpha (\lambda )},\;\delta \eta
_{(\lambda )}^{*}=\partial _{\alpha }\eta _{\;\;\;(\lambda )}^{*\alpha },
\label{cin12}
\end{equation}
\begin{equation}
\gamma B^{\alpha \beta (\lambda )}=\varepsilon ^{\alpha \beta \gamma \delta
}\partial _{\gamma }\eta _{\delta }^{\;\;(\lambda )},\;\gamma A_{\alpha
(\lambda )}=\partial _{\alpha }C_{(\lambda )},\;\gamma \eta _{\alpha
}^{\;\;(\lambda )}=\partial _{\alpha }\eta ^{(\lambda )},  \label{cin13}
\end{equation}
\begin{equation}
\gamma C_{(\lambda )}=\gamma \eta ^{(\lambda )}=0,\;\gamma B_{\alpha \beta
(\lambda )}^{*}=\gamma A^{*\alpha (\lambda )}=0,  \label{cin14}
\end{equation}
\begin{equation}
\gamma \eta _{\;\;\;(\lambda )}^{*\alpha }=\gamma C^{*(\lambda )}=\gamma
\eta _{(\lambda )}^{*}=0.  \label{cin15}
\end{equation}
The overall degree from the BRST complex is the ghost number ($\mathrm{gh}$%
), defined like the difference between the pure ghost number and the
antighost number, such that $\mathrm{gh}\left( s\right) =1$. The BRST
symmetry admits a canonical action in the antibracket $\left( ,\right) $, $%
sF=\left( F,S\right) $, where the canonical generator $S$ is bosonic, of
ghost number zero, and satisfies the classical master equation $\left(
S,S\right) =0$, which is equivalent to the second-order nilpotency of $s$, $%
s^{2}=0$. In the case of the free model under study, since both the gauge
generators and reducibility functions are field-independent, it follows that
the solution to the master equation is given by
\begin{equation}
S=S_{0}^{L}+\int d^{4}x\left( A^{*\alpha (\lambda )}\partial _{\alpha
}C_{(\lambda )}+\varepsilon ^{\alpha \beta \gamma \delta }B_{\alpha \beta
(\lambda )}^{*}\partial _{\gamma }\eta _{\delta }^{\;\;(\lambda )}+\eta
_{\;\;\;(\lambda )}^{*\alpha }\partial _{\alpha }\eta ^{(\lambda )}\right) .
\label{cin3}
\end{equation}

A consistent deformation of the free action (\ref{cin1}) and of its gauge
invariances (\ref{cin2}) defines a deformation of the corresponding solution
to the master equation that preserves both the master equation and the
field-ghost/antifield spectra. So, if $S_{0}^{L}+g\int d^{4}x\alpha
_{0}+O\left( g^{2}\right) $ stands for a consistent deformation of the free
action, with deformed gauge transformations $\bar{\delta}_{\epsilon
}A_{\alpha (\lambda )}=\partial _{\alpha }\epsilon _{(\lambda )}+g\beta
_{\alpha \lambda }+O\left( g^{2}\right) ,\;\bar{\delta}_{\epsilon }B^{\alpha
\beta (\lambda )}=\varepsilon ^{\alpha \beta \gamma \delta }\partial
_{\gamma }\epsilon _{\delta }^{\;\;(\lambda )}+g\beta ^{\alpha \beta \lambda
}+O\left( g^{2}\right) $, then the deformed solution to the master equation
\begin{equation}
\bar{S}=S+g\int d^{4}x\alpha +O\left( g^{2}\right) ,  \label{cin3a}
\end{equation}
satisfies $\left( \bar{S},\bar{S}\right) =0$, where the first-order
deformation $\alpha $ begins like $\alpha =\alpha _{0}+A^{*\alpha (\lambda )}%
\bar{\beta}_{\alpha \lambda }+B_{\alpha \beta (\lambda )}^{*}\bar{\beta}%
^{\alpha \beta \lambda }+`\mathrm{more}$' ($g$ is the so-called deformation
parameter or coupling constant). The terms $\bar{\beta}_{\alpha \lambda }$%
and $\bar{\beta}^{\alpha \beta \lambda }$ are obtained by replacing the
gauge parameters $\left( \epsilon _{(\lambda )},\epsilon _{\alpha
}^{\;\;(\lambda )}\right) $ respectively with the fermionic ghosts $\left(
C_{(\lambda )},\eta _{\alpha }^{\;\;(\lambda )}\right) $ in the functions $%
\beta _{\alpha \lambda }$ and $\beta ^{\alpha \beta \lambda }$.

The master equation $\left( \bar{S},\bar{S}\right) =0$ holds to order $g$ if
and only if
\begin{equation}
s\alpha =\partial _{\mu }j^{\mu },  \label{cin3b}
\end{equation}
for some local $j^{\mu }$. In order to solve this equation, we develop $%
\alpha $ according to the antighost number
\begin{equation}
\alpha =\alpha _{0}+\alpha _{1}+\cdots +\alpha _{I},\;\mathrm{agh}\left(
\alpha _{K}\right) =K,\;\mathrm{gh}\left( \alpha _{K}\right)
=0,\;\varepsilon \left( \alpha _{K}\right) =0.  \label{cin3c}
\end{equation}
The number of terms in the expansion (\ref{cin3c}) is finite and it can be
shown that we can take last term in $\alpha $ to be annihilated by $\gamma $%
, $\gamma \alpha _{I}=0$. Consequently, we need to compute the cohomology of
$\gamma $, $H\left( \gamma \right) $, in order to determine the component of
highest antighost number in $\alpha $. From (\ref{cin13}--\ref{cin15}) it is
simple to see that $H\left( \gamma \right) $ is spanned by $F_{\alpha \beta
(\lambda )}\equiv \partial _{\left[ \alpha \right. }A_{\left. \beta \right]
(\lambda )}$, $\partial _{\beta }B^{\beta \alpha (\lambda )}$ and $\chi
^{*}=\left( B_{\alpha \beta (\lambda )}^{*},A^{*\alpha (\lambda )},\eta
_{\;\;\;(\lambda )}^{*\alpha },C^{*(\lambda )},\eta _{(\lambda )}^{*}\right)
$, by their spacetime derivatives, as well as by the undifferentiated ghosts
$\left( C_{(\lambda )},\eta ^{(\lambda )}\right) $. (The derivatives of
these ghosts are removed from $H\left( \gamma \right) $ since they are $%
\gamma $-exact, in agreement with the second and third relations in (\ref
{cin13}).) If we denote by $e^{M}\left( C_{(\lambda )},\eta ^{(\lambda
)}\right) $ the elements with pure ghost number $M$ of a basis in the space
of the polynomials in the corresponding ghosts, it follows that the general
solution to the equation $\gamma \alpha _{I}=0$ takes the form
\begin{equation}
\alpha _{I}=a_{I}\left( \left[ F_{\alpha \beta (\lambda )}\right] ,\left[
\partial _{\beta }B^{\beta \alpha (\lambda )}\right] ,\left[ \chi
^{*}\right] \right) e^{I}\left( C_{(\lambda )},\eta ^{(\lambda )}\right) ,
\label{cin20}
\end{equation}
where $\mathrm{agh}\left( a_{I}\right) =I$. The notation $f\left( \left[
q\right] \right) $ means that $f$ depends on $q$ and its spacetime
derivatives up to a finite order. The equation (\ref{cin3b}) projected on
antighost number $\left( I-1\right) $ becomes
\begin{equation}
\delta \alpha _{I}+\gamma \alpha _{I-1}=\partial ^{\mu }\stackrel{(I-1)}{m}%
_{\mu }.  \label{cin21}
\end{equation}
Replacing (\ref{cin20}) in (\ref{cin21}), it follows that the last equation
possesses solutions with respect to $\alpha _{I-1}$ if the coefficients $%
a_{I}$ pertain to the homological space of the Koszul-Tate differential
modulo the exterior spacetime differential at antighost number $I$, $%
H_{I}\left( \delta |d\right) $, i.e., $\delta a_{I}=\partial _{\mu
}l_{I-1}^{\mu }$. In the meantime, since our free model is linear
and of Cauchy order equal to three, according to the results from
\cite{gen1} we have that $H_{J}\left( \delta |d\right) $ vanishes
for $J>3$, so we can assume that the first-order deformation stops
at antighost number three ($I=3$)
\begin{equation}
\alpha =\alpha _{0}+\alpha _{1}+\alpha _{2}+\alpha _{3},  \label{cin22}
\end{equation}
where $\alpha _{3}$ is of the form (\ref{cin20}), with $a_{3}$ from $%
H_{3}\left( \delta |d\right) $. On the one hand, the most general
representatives of $H_{3}(\delta |d)$ can be taken of the type
$\eta _{(\lambda )}^{*}$. On the other hand, the elements
$e^{3}\left( C_{(\lambda )},\eta ^{(\lambda )}\right) $ are
precisely $\left( \eta ^{(\mu )}C_{(\nu )},C_{(\mu )}C_{(\nu
)}C_{(\rho )}\right) $. Then, $\alpha _{3}$ is of the form $\eta
_{(\lambda )}^{*}\left( f_{\mu }^{\lambda \nu }\eta ^{(\mu
)}C_{(\nu )}+f^{\lambda \mu \nu \rho }C_{(\mu )}C_{(\nu )}C_{(\rho
)}\right) $, with $f_{\mu }^{\lambda \nu }$ and $f^{\lambda \mu
\nu \rho }$ some constants. By covariance arguments, $f_{\mu
}^{\lambda \nu }$ must contain at least one spacetime derivative,
which, if applied on the basis $\eta ^{(\mu )}C_{(\nu )}$, leads
to trivial ($\gamma $-exact) terms, such that we can set $f_{\mu
}^{\lambda \nu }=0$. In the meantime, the only manifestly
covariant constants $f^{\lambda \mu \nu \rho }$ in four spacetime
dimensions that do not involve spacetime derivatives can only be
proportional with the completely antisymmetric symbol $\varepsilon
^{\lambda \mu \nu \rho }$. In consequence, the last representative
from the first-order deformation (\ref {cin22}) reads as
\begin{equation}
\alpha _{3}=\frac{1}{3}\varepsilon ^{\lambda \mu \nu \rho }\eta _{(\lambda
)}^{*}C_{(\mu )}C_{(\nu )}C_{(\rho )}.  \label{cin25}
\end{equation}

By taking into account the relations (\ref{cin10}--\ref{cin15}), it follows
that the solution to the equation (\ref{cin21}) for $I=3$ is precisely given
by
\begin{equation}
\alpha _{2}=-\varepsilon ^{\lambda \mu \nu \rho }\left( \eta
_{\;\;\;(\lambda )}^{*\alpha }A_{\alpha (\mu )}+\varepsilon ^{\alpha \beta
\gamma \delta }\frac{1}{4}B_{\alpha \beta (\lambda )}^{*}B_{\gamma \delta
(\mu )}^{*}\right) C_{(\nu )}C_{(\rho )}+\xi C_{(\lambda )}^{*}\eta
^{(\lambda )},  \label{cin26}
\end{equation}
with $\xi $ a numerical constant. Further, we compute the component $\alpha
_{1}$ as solution to the equation $\delta \alpha _{2}+\gamma \alpha
_{1}=\partial ^{\mu }\stackrel{(1)}{m}_{\mu }$, and find that
\begin{eqnarray}
&&\alpha _{1}=B_{\alpha \beta (\lambda )}^{*}\left( \varepsilon
^{\alpha \beta \lambda \rho }+\varepsilon ^{\lambda \mu \nu \rho
}\varepsilon ^{\alpha \beta \gamma \delta }A_{\gamma (\mu
)}A_{\delta (\nu )}\right) C_{(\rho )}+ \nonumber \\
&&B_{\;\;\;\;(\nu )}^{*\mu \nu }C_{(\mu )}-\xi A_{\;\;\;(\lambda
)}^{*\alpha }\eta _{\alpha }^{\;\;(\lambda )}.  \label{cin27}
\end{eqnarray}
Finally, the antighost number zero piece in (\ref{cin22}) is subject to the
equation $\delta \alpha _{1}+\gamma \alpha _{0}=\partial ^{\mu }\stackrel{(0)%
}{m}_{\mu }$, whose solution can be written like
\begin{eqnarray}
&&\alpha _{0}=\varepsilon ^{\lambda \mu \nu \rho }\left(
A_{\lambda (\mu )}A_{\nu (\rho )}-\frac{1}{3!}\varepsilon ^{\alpha
\beta \gamma \delta }A_{\alpha (\lambda )}A_{\beta (\mu
)}A_{\gamma (\nu )}A_{\delta (\rho )}\right) + \nonumber \\
&&\frac{1}{2}\left( A_{\;\;(\mu )}^{\mu }A_{\;\;(\nu )}^{\nu
}-A^{\alpha (\mu )}A_{\mu (\alpha )}\right) -\frac{\xi
}{4}\varepsilon _{\alpha \beta \gamma \delta }B^{\alpha \beta
(\lambda )}B_{\;\;\;(\lambda )}^{\gamma \delta }. \label{cin28}
\end{eqnarray}
So far, we have completely generated the first-order deformation of the
solution to the master equation in the case of the analysed model, (\ref
{cin22}), where the concrete form of the terms $\left( \alpha _{a}\right)
_{a=0,1,2,3}$ can be found in the right hand-side of formulas (\ref{cin25}--%
\ref{cin28}).

Next, we investigate the equations that control the higher-order
deformations. If we denote by $S_{2}=\int d^{4}x\beta $ the second-order
deformation, the master equation $\left( \bar{S},\bar{S}\right) =0$ holds to
order $g^{2}$ if and only if
\begin{equation}
\frac{1}{2}\Delta =-s\beta +\partial _{\mu }t^{\mu },  \label{cin28a}
\end{equation}
where $\left( S_{1},S_{1}\right) =\int d^{4}x\Delta $. Making use
of (\ref {cin22}) and (\ref{cin25}--\ref{cin28}), we deduce that
\begin{eqnarray}
&&\frac{1}{2}\Delta =\partial _{\mu }t^{\mu }+\xi \left(
\varepsilon ^{\lambda \mu \nu \rho }\left( \frac{1}{3}C_{(\lambda
)}^{*}C_{(\rho )}+\eta _{(\lambda )}^{*}\eta _{(\rho )}+B_{\gamma
\delta (\lambda )}^{*}B_{\;\;\;(\rho )}^{\gamma \delta }\right)
C_{(\mu )}C_{(\nu )}+\right. \nonumber \\
&&2\left( \varepsilon ^{\lambda \mu \nu \rho }B_{\;\;\;(\lambda
)}^{\gamma \delta }A_{\gamma (\mu )}A_{\delta (\nu
)}+\frac{1}{4}\varepsilon ^{\gamma \delta \lambda \rho }B_{\gamma
\delta (\lambda )}-B_{\;\;\;(\lambda )}^{\rho \lambda }\right)
C_{(\rho )}- \nonumber \\
&&2\varepsilon ^{\lambda \mu \nu \rho }\left( \eta
_{\;\;\;(\lambda )}^{*\alpha }A_{\alpha (\mu
)}+\frac{1}{4}\varepsilon ^{\alpha \beta \gamma \delta }B_{\alpha
\beta (\lambda )}^{*}B_{\gamma \delta (\mu )}^{*}\right) C_{(\nu
)}\eta _{(\rho )}+A^{\alpha (\mu )}\eta _{\mu (\alpha )} \nonumber
\\
&&\left( B_{\alpha \beta (\lambda )}^{*}\left( \varepsilon
^{\alpha \beta \lambda \rho }+\varepsilon ^{\lambda \mu \nu \rho
}\varepsilon ^{\alpha \beta \gamma \delta }A_{\gamma (\mu
)}A_{\delta (\nu )}\right) +B_{\;\;\;\;(\nu )}^{*\rho \nu }\right)
\eta _{(\rho )}+ \nonumber \\
&&\varepsilon ^{\lambda \mu \nu \rho }\left( \left(
A_{\;\;\;(\lambda )}^{*\alpha }A_{\alpha (\mu )}+\eta
_{\;\;\;(\lambda )}^{*\alpha }\eta _{\alpha (\mu )}\right) C_{(\nu
)}+2\varepsilon ^{\alpha \beta \gamma \delta }B_{\alpha \beta
(\lambda )}^{*}A_{\gamma (\mu )}\eta _{\delta (\nu )}\right)
C_{(\rho )}+ \nonumber \\
&&\left. 2\left( \frac{1}{3}\varepsilon ^{\lambda \mu \nu \rho
}\varepsilon ^{\alpha \beta \gamma \delta }A_{\alpha (\lambda
)}A_{\beta (\mu )}A_{\gamma (\nu )}-\varepsilon ^{\alpha \mu
\delta \rho }A_{\alpha (\mu )}\right) \eta _{\delta (\rho
)}-A_{\;\;(\mu )}^{\mu }\eta _{\;\;(\nu )}^{\nu }\right) .
\label{cin29}
\end{eqnarray}
It is easy to see that none of the terms proportional with $\xi $ in the
right hand-side of (\ref{cin29}) can be written like an $s$-exact modulo $d$
quantity. In conclusion, the consistency of the first-order deformation
requires that $\xi =0$. With this value at hand, the equation (\ref{cin28a})
is satisfied with the choice $\beta =0$, which further induces that the
second-order deformation of the solution to the master equation can be taken
to vanish, $S_{2}=0$. Further, all the higher-order equations are satisfied
if we set $S_{3}=S_{4}=\cdots =0$. Consequently, the complete deformed
solution to the master equation, consistent to all orders in the coupling
constant, reduces in this situation to the sum between the free solution (%
\ref{cin3}) and the first-order deformation in which we replace
$\xi $ by zero, and hence is expressed by
\begin{eqnarray}
&&\bar{S}=S_{0}^{L}+\int d^{4}x\left( g\varepsilon ^{\lambda \mu
\nu \rho }\left( A_{\lambda (\mu )}A_{\nu (\rho
)}-\frac{1}{3!}\varepsilon ^{\alpha \beta \gamma \delta }A_{\alpha
(\lambda )}A_{\beta (\mu )}A_{\gamma (\nu )}A_{\delta (\rho
)}\right) +\right. \nonumber \\
&&\frac{g}{2}\left( A_{\;\;(\mu )}^{\mu }A_{\;\;(\nu )}^{\nu
}-A^{\alpha (\mu )}A_{\mu (\alpha )}\right) +A^{*\alpha (\lambda
)}\partial _{\alpha }C_{(\lambda )}+\varepsilon ^{\alpha \beta
\gamma \delta }B_{\alpha \beta (\lambda )}^{*}\partial _{\gamma
}\eta _{\delta }^{\;\;(\lambda )}+ \nonumber \\
&&gB_{\alpha \beta (\lambda )}^{*}\left( \frac{1}{2}\left(
g^{\alpha \rho }g^{\beta \lambda }-g^{\alpha \lambda }g^{\beta
\rho }\right) +\varepsilon ^{\alpha \beta \lambda \rho
}+\varepsilon ^{\lambda \mu \nu \rho }\varepsilon ^{\alpha \beta
\gamma \delta }A_{\gamma (\mu )}A_{\delta (\nu )}\right) C_{(\rho
)}+ \nonumber \\
&&\eta _{\;\;\;(\lambda )}^{*\alpha }\partial _{\alpha }\eta
^{(\lambda )}-g\varepsilon ^{\lambda \mu \nu \rho }\left( \eta
_{\;\;\;(\lambda
)}^{*\alpha }A_{\alpha (\mu )}+\varepsilon ^{\alpha \beta \gamma \delta }%
\frac{1}{4}B_{\alpha \beta (\lambda )}^{*}B_{\gamma \delta (\mu
)}^{*}\right) C_{(\nu )}C_{(\rho )}+ \nonumber \\
&&\left. \frac{g}{3}\varepsilon ^{\lambda \mu \nu \rho }\eta
_{(\lambda )}^{*}C_{(\mu )}C_{(\nu )}C_{(\rho )}\right) .
\label{cindef}
\end{eqnarray}

With the help of the last formula, we are able to identify the interacting
tensor gauge field theory behind the deformation procedure. For instance,
the antighost number zero component in (\ref{cindef}) is nothing but the
Lagrangian action of the resulting coupled model
\begin{eqnarray}
&&\bar{S}_{0}^{L}\left[ A_{\alpha (\lambda )},B^{\alpha \beta
(\lambda )}\right] =\int d^{4}x\left( \partial _{\left[ \alpha
\right. }A_{\left. \beta \right] (\lambda )}B^{\alpha \beta
(\lambda )}+g\varepsilon ^{\lambda \mu \nu \rho }A_{\lambda (\mu
)}A_{\nu (\rho )}+\right. \nonumber \\
&&\left. \frac{g}{2}\left( A_{\;\;(\mu )}^{\mu }A_{\;\;(\nu
)}^{\nu }-A^{\alpha (\mu )}A_{\mu (\alpha )}\right)
-\frac{g}{3!}\varepsilon ^{\lambda \mu \nu \rho }\varepsilon
^{\alpha \beta \gamma \delta }A_{\alpha (\lambda )}A_{\beta (\mu
)}A_{\gamma (\nu )}A_{\delta (\rho )}\right) . \label{cin30}
\end{eqnarray}
From the elements of antighost number one in (\ref{cindef}), we
read the deformed gauge transformations of the tensor fields
\begin{eqnarray}
&&\bar{\delta}_{\epsilon }A_{\alpha (\lambda )}=\partial _{\alpha
}\epsilon _{(\lambda )}\equiv \left( Z_{\alpha \lambda
}^{(A)}\right) ^{\rho }\epsilon _{(\rho
)},\;\bar{\delta}_{\epsilon }B^{\alpha \beta (\lambda
)}=\varepsilon ^{\alpha \beta \gamma \delta }\partial _{\gamma
}\epsilon _{\delta }^{\;\;(\lambda )}+\frac{g}{2}\epsilon ^{\left[
(\alpha )\right. }g^{\left. \beta \right] \lambda }+ \nonumber \\
&&g\left( \varepsilon ^{\alpha \beta \lambda \rho }+\varepsilon
^{\lambda \mu \nu \rho }\varepsilon ^{\alpha \beta \gamma \delta
}A_{\gamma (\mu )}A_{\delta (\nu )}\right) \epsilon _{(\rho
)}\equiv \nonumber \\
&&\left( Z^{(B)\alpha \beta \lambda }\right)
_{\tau }^{\delta }\epsilon _{\delta }^{\;\;(\tau )}+\left(
Z^{(B)\alpha \beta \lambda }\right) ^{\rho }\epsilon _{(\rho )},
\label{cin31}
\end{eqnarray}
where the nonvanishing gauge generators involved with the coupled theory are
given by
\begin{equation}
\left( Z_{\alpha \lambda }^{(A)}\right) ^{\rho }=\delta _{\lambda
}^{\;\;\rho }\partial _{\alpha },\;\left( Z^{(B)\alpha \beta \lambda
}\right) _{\tau }^{\delta }=\varepsilon ^{\alpha \beta \gamma \delta }\delta
_{\;\;\tau }^{\lambda }\partial _{\gamma },  \label{cin34}
\end{equation}
\begin{equation}
\left( Z^{(B)\alpha \beta \lambda }\right) ^{\rho }=\frac{g}{2}\left(
g^{\alpha \rho }g^{\beta \lambda }-g^{\alpha \lambda }g^{\beta \rho }\right)
+g\left( \varepsilon ^{\alpha \beta \lambda \rho }+\varepsilon ^{\lambda \mu
\nu \rho }\varepsilon ^{\alpha \beta \gamma \delta }A_{\gamma (\mu
)}A_{\delta (\nu )}\right) .  \label{cin33}
\end{equation}
Related to the antighost number two contribution of (\ref{cindef}), we
remark that the reducibility functions and relations are not modified during
the deformation mechanism with respect to the initial free model. In change,
the original abelian gauge algebra is deformed into an open one, where the
associated non-abelian commutators among the gauge generators are provided
by
\begin{eqnarray}
&&\left( Z_{\delta \tau }^{(A)}\right) ^{\sigma }\frac{\delta
\left(
Z^{(B)\alpha \beta \lambda }\right) ^{\rho }}{\delta A_{\delta (\tau )}}%
-\left( Z_{\delta \tau }^{(A)}\right) ^{\rho }\frac{\delta \left(
Z^{(B)\alpha \beta \lambda }\right) ^{\sigma }}{\delta A_{\delta
(\tau )}}= \nonumber \\
&&-2g\varepsilon ^{\sigma \rho \tau \mu }A_{\delta (\mu )}\left(
Z^{(B)\alpha \beta \lambda }\right) _{\tau }^{\delta
}+g\varepsilon ^{\sigma \rho \lambda \mu }\varepsilon ^{\alpha
\beta \gamma \delta } \frac{\delta \bar{S}_{0}^{L}}{\delta
B^{\gamma \delta (\mu )}}.  \label{cin36}
\end{eqnarray}
The pieces of antighost number three in (\ref{cindef}) offer information on
the second-order structure functions due to the open character of the
deformed gauge algebra.

In conclusion, in this paper we have investigated a special class
of mixed-symmetry type tensor gauge fields of degrees two and
three in four dimensions from the perspective of the Lagrangian
deformation procedure based on cohomological BRST techniques. We
have shown that the deformed solution to the master equation can
be taken to be nonvanishing only at the first order in the
coupling constant. Thus, we reveal an interacting model with: (i)
deformed gauge transformations; (ii) an open gauge algebra; (iii)
a quartic vertex and ``mass'' terms involving only the tensors of
degree two; (iv) undeformed reducibility functions. It is
interesting to mention that, in spite of the fact that the tensor
gauge fields involved here are of the same type with those studied
in \cite{EPJC} (up to the notational replacement
$A\longleftrightarrow B$), the resulting interacting theories are
in a way complementary. More precisely, while here we deform the
gauge algebra, but do not affect the reducibility, in \cite{EPJC}
the reducibility is essentially changed, although the gauge
algebra is not modified. Moreover, the first-order deformation of
the solution to the master equation derived here contains no
spacetime derivatives, while that from \cite{EPJC} involves only
derivative terms. Finally, we note that those interactions from
\cite{EPJC} that actually deform the gauge symmetry only exist in
four spacetime dimensions, by contrast to the model discussed
here, which is suited to further generalization. In a future paper
we hope to solve this problem.

\section*{Acknowledgments}

Some of the authors (C.B., S.O.S. and S.C.S.) acknowledge
financial support from a type-A grant with CNCSIS-MEC, Romania.

\end{document}